\documentclass{PoS}

\title{EXO 2030+375 Restarts in Reverse}

\ShortTitle{EXO 2030+375 Restarts in Reverse}

\author{\speaker{Peter Kretschmar}\thanks{We thank the Swift, NuSTAR, and NOT teams for scheduling the requested observations, sometimes on short notice.}\\
        European Space Astronomy Centre (ESA/ESAC), Science Operations Department\\
        P.O.~Box 78, E-28691, Villanueva de la Ca\~{n}ada, Madrid, Spain\\
        E-mail: \email{peter.kretschmar@esa.int}}

\author{Felix Fuerst\\
        Cahill Center for Astronomy and Astrophysics, California Institute of Technology, Pasadena, CA
91125, USA}
        
\author{Colleen A. Wilson-Hodge\\     
        NASA Marshall Space Flight Center, Huntsville, AL 35812, USA}

\author{Pere Blay\\
        Nordic Optical Telescope -- IAC, P.O.Box 474, E-38700, Santa Cruz de La Palma\\
Santa Cruz de Tenerife, Spain}
        
\author{Jari Kajava\\
        European Space Astronomy Centre (ESA/ESAC), Science Operations Department\\
        P.O.~Box 78, E-28691, Villanueva de la Ca\~{n}ada, Madrid, Spain}
        
\author{Julia Alfonso-Garz{\'o}n\\
Centro de Astrobiolog{\'\i}a (CSIC--INTA), Camino Bajo del Castillo s/n,\\ 
Urb. Villafranca del Castillo, E-28692 Villanueva de la Ca{\~n}ada, Madrid, Spain}
        
\author{Matthias K{\"u}hnel\\
        Dr. Karl Remeis-Observatory \& ECAP, Universit{\"a}t Erlangen-N{\"u}rnberg, \\
        Sternwartstr. 7, D-96049 Bamberg, Germany}
        
\author{Ingo Kreykenbohm\\
        Dr. Karl Remeis-Observatory \& ECAP, Universit{\"a}t Erlangen-N{\"u}rnberg, \\
        Sternwartstr. 7, D-96049 Bamberg, Germany}
        
\author{J\"orn Wilms\\
        Dr. Karl Remeis-Observatory \& ECAP, Universit{\"a}t Erlangen-N{\"u}rnberg, \\
        Sternwartstr. 7, D-96049 Bamberg, Germany}        

\author{Peter A. Jenke\\     
        CSPAR, SPA University of Alabama in Huntsville, Huntsville, AL 35805, USA}

\author{Katja Pottschmidt\\     
        Department of Physics \& Center for Space Science and Technology, University of Maryland Baltimore County, Baltimore, MD 21250, USA\\
        CRESST \& NASA Goddard Space Flight Center, Code 661, Greenbelt, MD 20771, USA}
        
\abstract{The Be X-ray binary pulsar EXO 2030+375, first detected in 1985, 
has shown a significant detected X-ray outburst at nearly every periastron 
passage of its 46-day orbit for the past ~25 years, with one low state
accompanied by a torque reversal in the 1990s. In early 2015 the 
outbursts progressively became fainter and less regular while the monotonic 
spin-up flattened.
At the same time a decrease in the H$\alpha$ line equivalent width was
reported, indicating a change in the disk surrounding the mass donor.

In order to explore the source behaviour in the
poorly explored low-flux state with a possible transition to a state of
centrifugal inhibition of accretion we have undertaken an observing
campaign with Swift/XRT, NuSTAR and the Nordic Optical Telescope (NOT).
This conference contribution reports the preliminary results obtained
from our campaign.
}

\FullConference{11th INTEGRAL Conference Gamma-Ray Astrophysics in Multi-Wavelength Perspective,\\
		10-14 October 2016\\
		Amsterdam, The Netherlands}

\begin{document}
\section{Introduction}
EXO 2030+375 is a well-known transient Be X-ray Binary, discovered during a giant outburst in 1985 
\cite{IAUC4066,Parmar:89a}. Since 1991 normal (type~I) outbursts have been observed at nearly every periastron passage of its 46 day orbit. From 1992 to 1994 outbursts were bright and the pulsar was spinning up. Then the flux levels dropped suddenly and a global spin-down trend ensued. Furthermore, 1995 the orbital phase of the type~I outbursts shifted to 8--9 days earlier \cite{ReigCoe:98,Wilson:2002}.

\begin{figure} 
\centerline{\includegraphics[width=1.0\textwidth]{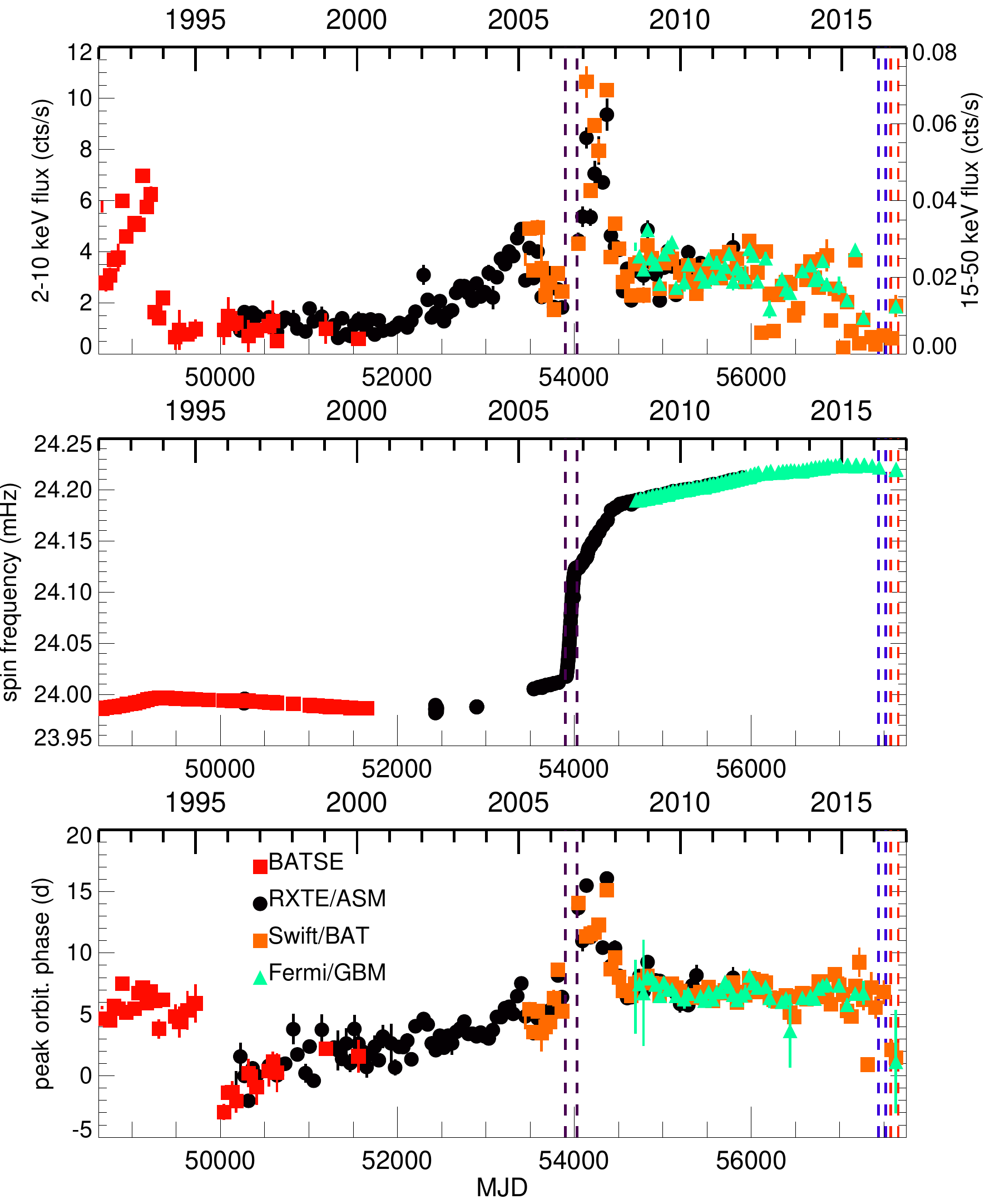}}
\caption{Long-term history of outburst peak fluxes, spin frequency and orbital phase of outburst peaks in EXO 2030+375. Note that the giant outburst of 2006 is not depicted in the top and bottom panels, as it is off-scale in flux ($\sim$40 cts/s in the 5--10 keV, $\sim$0.23 cts/s in the 15--50 keV band) and went on for multiple orbital periods. The rough timerange of this outburst is marked by dashed purple lines. At the right hand of the plots, other dashed lines mark the periods covered by our TOO observations with Swift and NuSTAR.} \label{fig:long-term} 
\end{figure}

Since 2002 the outbursts brightened and the spin trend reversed again. Another giant outburst took place in 
summer 2006, accelerating the pulsar significantly \cite{Wilson:2008}. For more than eight years regular 
outbursts and the spin-up trend, monitored first by RXTE and then the Fermi/GBM, continued 
(see Figure~\ref{fig:long-term}), albeit gradually reducing in amplitude.

In early 2015 the regularity of outbursts decreased with first every second outburst much fainter than 
usual and then hardly any outburst activity at all. At the same time, the spin-up trend flattened to almost 
zero \cite{ATel8835}. 
Optical spectroscopic observations in this period demonstrated a clear decrease in the H$\alpha$ emission equivalent width, which is usually taken as a measure of the size of the disk around the Be star feeding regular outbursts.

\begin{figure} 
\centerline{\includegraphics[width=.8\textwidth]{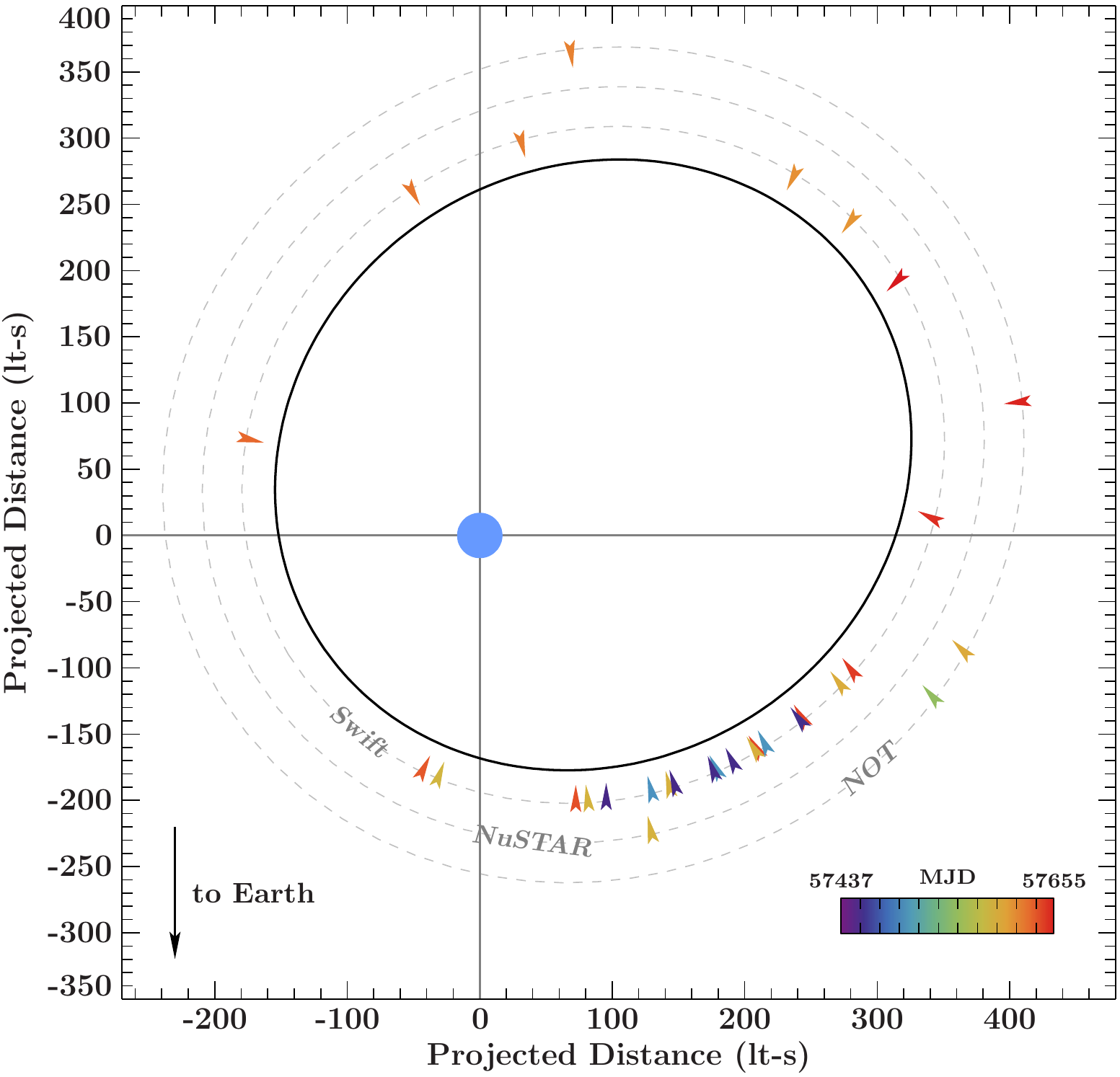}}
\caption{Sketch of the EXO 2030+375 system with markers showing the phasing of the different observations
of our monitoring program. Marker colours indicate the time of the respective observations.} \label{fig:coverage} 
\end{figure}

\section{Observations}
Prompted by the evident change in source behaviour, we began to observe
EXO~2030+375 with Swift/XRT. First with a few snapshots, then with an extended monitoring,
with a tighter monitoring around the expected outburst peaks and increasingly wider spaced observations 
around the rest of the orbit. In addition, we obtained one 60~ks NuSTAR observation and took optical spectra 
of the H$\alpha$ line with the Nordic Optical Telescope on four occasions  of our campaign. The orbital
phases of our observations are indicated in Figure~\ref{fig:coverage}.


\section{Preliminary Results}
In our targeted X-ray observations, we always find some low-level pulsed emission from EXO 2030+375, thus no sign of a magnetospheric cutoff of emission (a.k.a.\ ``propeller effect''). The luminosity during the faint states is low enough that accretion directly from the stellar wind can account for it.
While the absorption and spectral hardness are rather constant despite strong flux variations, we find a much higher absorption in one observation during the last monitored outburst, possibly the neutron star passing behind the Be star disk.
\begin{figure} 
\centerline{\includegraphics[width=1.0\textwidth]{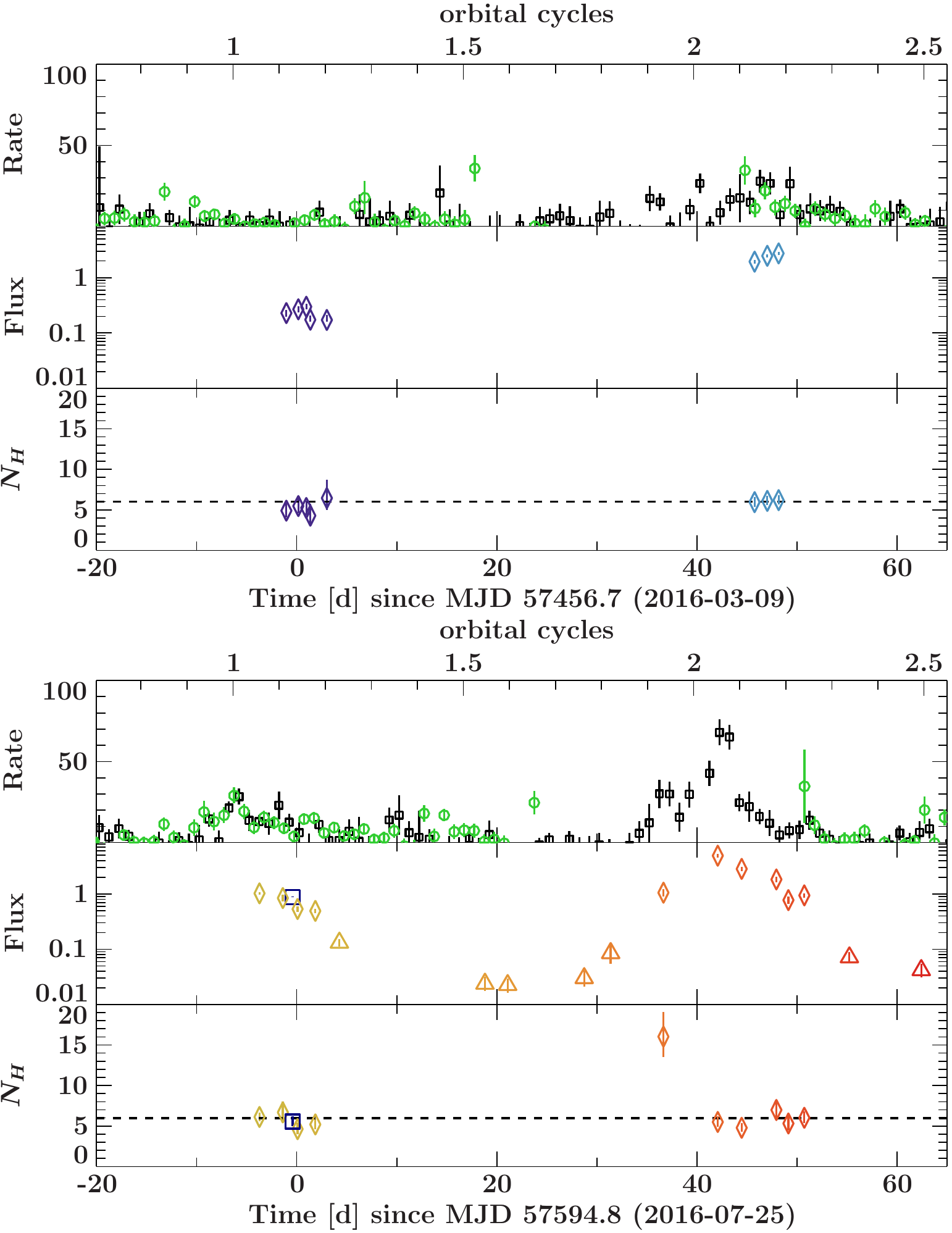}}
\caption{X-ray observations. The top panel shows fluxes determined by the MAXI (green) and Swift/BAT (black) monitoring instruments. The flux and N$_\mathrm{H}$ values have been obtained from spectral fits to Swift/XRT (diamonds and triangles) and NuSTAR data. Triangles indicate observations with low countrate statistics for which the N$_\mathrm{H}$ value was fixed to 5.3, the average value indicated by the dashed horizontal line. Note the logarithmic scale of the middle panel.} \label{fig:results} 
\end{figure}

Both Fermi/GBM and Swift/XRT clearly show that EXO 2030+375 is again in spin-down. This torque reversal is 
remarkably similar in duration and magnitude to the one observed $\sim$21 years ago, as is the shift in the 
orbital phase of outburst peaks. The same time interval lies between the two known giant outbursts, 
suggesting a $\sim$21 year quasi-period in this system, as also noted by \cite{ATel9263,Laplace:2016X}.

The minimum EW we measure is similar to the values measured by \cite{Wilson:2002,Reig:98}
after the spin-down and orbital phase shift of 1995, and by \cite{Baykal:2008} after
the giant outburst, pointing to a relation between the circumstellar
state disc and the changes in the pulsar behaviour.

The H$\alpha$ profile clearly evolves between the first two observations (June 15$^\mathrm{th}$, MJD 57554 
and August 1$^\mathrm{st}$, MJD 57601) -- indicating an important change in structure, density gradient,
size or geometry of the Be disk -- but then remains similar for the remaining observations.

\clearpage
\begin{figure} 
\centerline{\includegraphics[width=0.8\textwidth]{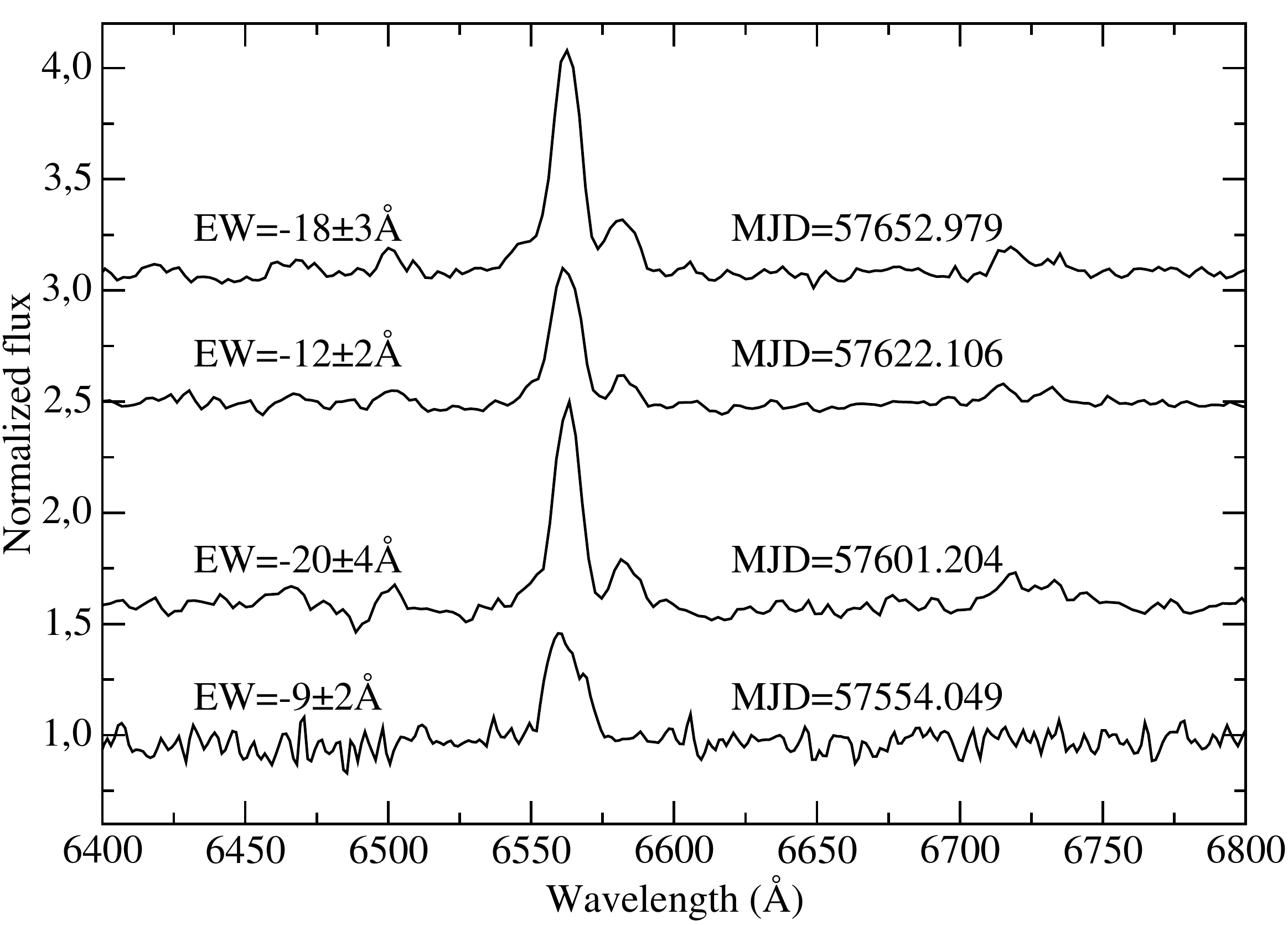}}
\caption{H$\alpha$ spectra of V$^\star$ V2246~Cyg taken at the NOT telescope between June and September 2016} \label{fig:halpha} 
\end{figure}

\end{document}